\documentclass{ws-ijmpa}
\usepackage{graphicx}

\begin{document}

\markboth{Thiago Prudencio}
{Parametric down conversion of a bosonic thermofield vacuum}
\catchline{}{}{}{}{}

\title{PARAMETRIC DOWN CONVERSION OF A BOSONIC THERMOFIELD VACUUM}

\author{Thiago Prud\^encio}

\address{Instituto de F\'\i sica, Universidade de Bras\'ilia - UnB, CP: 04455, 70919-970, Bras\'ilia - DF, Brazil. \\
International Institute of Physics, Universidade Federal do
Rio Grande do Norte, Av. Odilon Gomes de Lima, 1722, 59078--400, Natal, RN, Brazil.
\footnote{thprudencio@gmail.com}}

\maketitle

\begin{abstract}
We consider a process of parametric down conversion where the input state is a bosonic
thermofield vacuum. This state leads to a parametric down conversion, generating an
output of two excited photons. Following a thermofield dynamics scheme, the input state,
initially in a bosonic thermofield vacuum, and the output states, initially in vacuum
states, evolve under a Liouville-von Neumann equation.
\keywords{down conversion; vacuum; thermofield}
\end{abstract}

\ccode{PACS numbers: 42.50.-p, 42.65.Yj, 78.20.N-}

\section{Introduction}

Proposed in 1975, by Takahashi and Umezawa\cite{takahashi}, thermofield dynamics (TFD) is 
an operator-algebraic approach to quantum statistical mechanics and a real time formalism to 
finite temperature quantum field theory\cite{landsman,matsumoto2,das,umezawab,umezawab2,santanab}. The basic ingredients of 
TFD are the doubling of freedom degrees in the Hilbert space $\mathcal{H}$ which describes 
the physical system and the building of a finite temperature vacuum, thermofield vacuum, by means of 
a Bogoliubov transformation realized in the zero temperature vacuum state defined into the 
Hilbert space $\mathcal{H}\otimes\tilde{\mathcal{H}}$, also called 
Liouville space \cite{ban}. This procedure is made in such 
a way that the expectation value of any operator from $\mathcal{H}$ in the 
thermofield vacuum coincides with the statistical mean value. 

TFD has been largely applied in the study of finite temperature systems, high 
energy physics \cite{kobes,costa,leineker,balachandran,ojima,nakano1,silva,belich,leblanc,nakahara}, quantum optics 
\cite{knight,barnett,mann,chaturvedi,vourdas,matrasulov,tay}, condensed matter physics \cite{suzuki,matsumoto4},
among others\cite{yamanaka,umezawab,umezawab2,santanab}.

Here we consider a process where the input state is a bosonic thermofield vacuum. This state will be lead to a parametric 
down conversion interaction generating photon excitations as output states. In general, this process 
is described by a way where the input state
suffering the parametric down conversion is completelly annihilated generating as output correlated photons with
number operator expectation values equal to 1, where the output states were 
initially vacuum state. 

Indeed, parametric down conversion is a non-linear process in the light of quantizing electromagnetic field where
a single photon incides on a crystal with second order non-linearity generating in the interaction two photons with resulting
frequencies equal to the sum of the input frequency. The output frequencies are called generally signal and idler, the
 frequency of the input photon is called generally pump frequency \cite{vedral,walls}.
 
In the process that we will describe, 
our aim is to obtain as output states one-photon number excitations, considering the distribution
of the photon number operator. Following a TFD scheme where the system evolves under a Liouville-von Neumann equation, 
the input state, initially in a bosonic thermofield vacuum, and the output states, initially in vacuum states, are changed
under the parametric down conversion leading the output states one-photon excitations whose caracterizations is described
by the mean expectation value of their number operators.  

In fact, thermofield dynamics have been used for studying parametric amplification \cite{knight} and 
degenerate parametric amplification with dissipation \cite{arimitsu}. This last can treat the case where the
bosonic thermofield vacuum evolves under dissipation, such that part of the hamiltonian is non-hermitean. 

In our propose, a parametric down conversion of a bosonic thermofield vacuum is considered 
 using equilibrium states, such 
that the Liouville-von Neumann operator can be used in an hermitean form.    

\section{Thermofield vacuum}

Giving an operator $\hat{A}$ acting on a Hilbert space $\mathcal{H}$ generated by Fock states $|n\rangle$, its expectation value in a given ensemble is expressed by
\begin{eqnarray}
\langle \hat{A}\rangle = Tr(\hat{\rho} \hat{A}), 
\end{eqnarray}
where $\hat{\rho}$ is the density operator in the corresponding ensemble. In thermofield dynamics this expectation value is evaluated by means of the definition of a thermofield vacuum $
|0(\beta )\rangle $, where $\beta=1/T$ is the inverse of temperature $T$ ($k_{B}=\hbar=1$), leading to the same result as statistical approach, i.e.,
\begin{eqnarray}
\langle 0(\beta )|\hat{A}|0(\beta )\rangle = Tr(\hat{\rho} \hat{A}). 
\label{med}
\end{eqnarray}
As a consequence, the thermal vaccuum state $|0(\beta )\rangle$ is associated to the density
operator $\hat{\rho}$. For this reason we need to describe it in a Hilbert space larger than the Hilbert space $\mathcal{H}$ generated by the Fock states $|n\rangle$. Then, the thermofield vacuum $|0(\beta )\rangle$ is not a vector state in the Hilbert space $\mathcal{H}$ described by the Fock states $|n\rangle$, but a state in another enlarged Hilbert space $\mathcal{H}\otimes\tilde{\mathcal{H}}$, where $\tilde{\mathcal{H}}$ is the Hilbert space conjugated to $\mathcal{H}$. In fact, in order to describe $|0(\beta)\rangle $ as a vector state, we need to double the degrees of freedom of the Hilbert space $\mathcal{H}$ by a formal procedure named tilde conjugation \cite{santanab}, creating the space $\mathcal{H}\otimes\tilde{\mathcal{H}}$.

Consider the system described by a thermal equilibrium density matrix
\begin{eqnarray}
\hat{\rho}=e^{-\beta \hat{H}}/Z,
\end{eqnarray}
where $Z=Tr(e^{-\beta \hat{H}})$ is the partition function and the energy spectrum $E_{n}$ of the hamiltonian $\hat{H}$ is given by
$\hat{H}|n\rangle = E_{n}|n\rangle$.
We construct the thermofield vacuum in the space $\mathcal{H}\otimes\tilde{\mathcal{H}}$, 
in terms of the Fock state basis $|n\rangle$ and unknown vectors $|c_{n} \rangle \in \tilde{\mathcal{H}}$,
 by means of the following expansion
\begin{eqnarray}
|0(\beta)\rangle =\sum_{n}|c_{n} \rangle|n\rangle.
\label{evt}
\end{eqnarray}
In order to find the vectors $|c_{n}\rangle$, we can use the equation (\ref{med}),
where the operator $\hat{A}$ does not act on $|c_{n} \rangle$ vectors,
\begin{eqnarray}
\langle 0(\beta )|\hat{A}|0(\beta )\rangle =\sum_{m,n}\langle c_{m}|c_{n} \rangle\langle m|\hat{A}|n\rangle.
\label{somado}
\end{eqnarray}
On the other hand, we can write
\begin{eqnarray}
Tr(\hat{\rho} \hat{A})=\frac{1}{Z}\sum_{n}e^{-\beta E_{n}}\langle n|\hat{A}|n\rangle.
\label{rodado}
\end{eqnarray}
Comparing equations (\ref{rodado}) and (\ref{somado}), we find
\begin{eqnarray}
\langle c_{m}|c_{n} \rangle = \frac{1}{Z}e^{-\frac{1}{2}\beta(E_{m} + E_{n})}\delta_{mn}.
\end{eqnarray}
Thus, by defining the Fock states $|\tilde{n}\rangle \in \tilde{\mathcal{H}}$ a basis product can be given to $\mathcal{H}\otimes\tilde{\mathcal{H}}$ and we can write $|c_{n} \rangle$ vectors as
\begin{equation}
|c_{n} \rangle = \frac{1}{\sqrt{Z}}e^{-\frac{1}{2}\beta E_{n}}|\tilde{n}\rangle.
\label{vec1}
\end{equation}
Now, we can write the thermofield vacuum in terms of $|n,\tilde{n}\rangle \in \mathcal{H}\otimes\tilde{\mathcal{H}}$,
\begin{eqnarray}
|0(\beta )\rangle =\frac{1}{\sqrt{Z}}\sum_{n}e^{-\beta E_{n}/2}|n,\tilde{n}\rangle.
\end{eqnarray}
In fact, the term vacuum is only appropriated because we can define annihilation and creation thermofield operators, 
that commute in the bosonic case and anti-commute in fermionic case,
\begin{eqnarray}
\bar{a}_{\beta}^{\dagger } &=& e^{-i\bar{G}}\hat{a}^{\dagger }e^{i\bar{G}}, \label{lyiul1}\\
\bar{a}_{\beta} &=& e^{-i\bar{G}}\hat{a}e^{i\bar{G}} \label{lyiul2},\\
\bar{b}_{\beta}^{\dagger } &=& e^{-i\bar{G}}\tilde{b}^{\dagger }e^{i\bar{G}}, \label{lyiul3}\\
\bar{b}_{\beta} &=& e^{-i\bar{G}}\tilde{b}e^{i\bar{G}}. \label{lyiul4}
\end{eqnarray}
and the thermofield vacuum can be annihilated
\begin{eqnarray}
\bar{a}_{\beta}|0(\beta)\rangle = \bar{b}_{\beta}|0(\beta)\rangle=0,
\label{671878}
\end{eqnarray}
where $\bar{G}$ is an unitary operator mixing $|n\rangle \in \mathcal{H}$ and $|\tilde{n}\rangle \in \tilde{\mathcal{H}}$ 
by acting on the Hilbert space $\mathcal{H}\otimes\tilde{\mathcal{H}}$, as a two-mode squeezing operator,
\begin{eqnarray}
\bar{G}=i\theta(\beta)\left(\hat{a}^{\dagger}\tilde{b}^{\dagger} - \hat{a}\tilde{b} \right),
\label{bog}
\end{eqnarray}
where $\theta=\theta(\beta)$ is a parameter related to a thermal distribution, $\hat{a}^{\dagger}$ and $\tilde{b}^{\dagger}$ 
are creation operators acting on spaces $\mathcal{H}$ and $\tilde{\mathcal{H}}$, and 
$\hat{a}$ and $\tilde{b}$ are annihilation operators acting on spaces $\mathcal{H}$ and $\tilde{\mathcal{H}}$, respectively. 
In the case of a bosonic oscillator system, we have $\theta=\tanh^{-1}(e^{-\beta\omega/2})$, related to a Bose-Einsten statistics, 
and for a fermionic oscillator system we have $\theta=\tan^{-1}(e^{-\beta\omega/2})$, 
related to a Fermi-Dirac statistics. The term $e^{-i\bar{G}}$ is a Bogoliubov transformation.
Since the Bogoliubov transformation is canonical, the corresponding commutations for bosons or anticommutations for fermions are preserved.

The operators $\bar{a}_{\beta}^{\dagger }$ and $\bar{b}_{\beta}^{\dagger }$, given by equations 
(\ref{lyiul1}) and (\ref{lyiul3}), excite the thermofield vacuum generating excited thermofield states.

In the Liouville space $\mathcal{H}\otimes\tilde{\mathcal{H}}$ given by TFD, the zero temperature vacuum state is given by $|0,\tilde{0}\rangle $, associated to a density operator at zero temperature $\hat{\rho}_{0} =|0\rangle \langle 0| \in \mathcal{H}$.
By applying a Bogoliubov transformation $e^{-i\bar{G}}$ on this vacuum state $|0,\tilde{0}\rangle$, the thermofield vacuum is generated at a finite temperature $T=\beta^{-1}$,
\begin{eqnarray}
|0(\beta )\rangle =e^{-i\bar{G}}|0,\tilde{0}\rangle.
\label{adk}
\end{eqnarray}

In TFD, tilde conjugation rules realize a mapping between $\hat{A}$ operators acting on $|n\rangle$
 and $\tilde{A}$ acting on $|\tilde{n}\rangle$. These rules are summarized by
\begin{eqnarray}
\widetilde{(\hat{A}\hat{B})}&=&\tilde{A}\tilde{B},\\
\widetilde{(z\hat{A}+w\hat{B})}&=& z^{\ast }\tilde{A}+w^{\ast }\tilde{B},\\
\widetilde{(\hat{A}^{\dagger })}&=&(\tilde{A})^{\dagger },\\
\widetilde{(\tilde{A})}&=& \pm\hat{A}, \label{pm}\\
\lbrack \hat{A},\tilde{B}]_{\pm}&=&0, \label{pm2}
\end{eqnarray}
where the operators $\hat{A}$ and $\hat{B}$ act only in the Hilbert space spanned
by $|n\rangle$, and $\tilde{A}$ and $\tilde{B}$ act only in the Hilbert
space generated by $|\tilde{n}\rangle $, where $z$ and $w$ are complex
numbers, $z^{\ast }$ and $w^{\ast }$ are their respective complex conjugated. 
In equation (\ref{pm}), $+$ is for bosons and $-$ is for fermions \cite{ojima}. 
In equation (\ref{pm2}) $+$ means commutation for bosons and $-$ is anticommutation for fermions. 

\begin{figure}[h]
\centering
\includegraphics[scale=0.4]{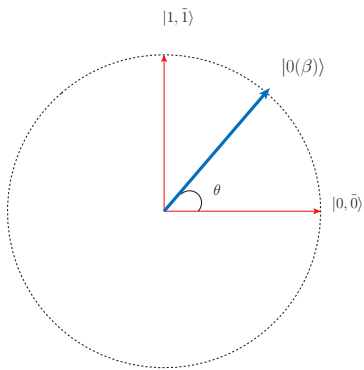}
\caption{(Color online) Fermionic thermofield vacuum in the plane generated by $|0,\tilde{0}\rangle$ and $|1,\tilde{1}\rangle$.}
\label{theo12}
\end{figure}

In the case of a fermionic oscillator, the space $\mathcal{H}\otimes\tilde{\mathcal{H}}$ is generated 
from the zero temperature vacuum and its excitations,
\begin{eqnarray}
|0,\tilde{0}\rangle &=& 1|0,\tilde{0}\rangle,  \label{00}\\
|1,\tilde{0}\rangle &=& \hat{a}^{\dagger }|0,\tilde{0}\rangle, \label{10} \\
|0,\tilde{1}\rangle &=&\tilde{b}^{\dagger }|0,\tilde{0}\rangle, \label{01} \\
|1,\tilde{1}\rangle &=& \hat{a}^{\dagger }\tilde{b}^{\dagger }|0,\tilde{0}\rangle. \label{11}
\end{eqnarray}
By applying the Bogolioubov transformation $e^{-i\bar{G}}$ on the vacuum $|0,\tilde{0}\rangle$
 we arrive at the following fermionic thermofield vacuum
\begin{eqnarray}
|0(\beta )\rangle = \frac{1}{\sqrt{Z}}(|0,\tilde{0}\rangle +e^{\frac{%
-\beta \omega }{2}}|1,\tilde{1}\rangle).
\end{eqnarray}
From the normalization condition $|0(\beta)\rangle $, we derive a partition
function $Z=1+e^{-\beta \omega }$. This state can also be written as
\begin{equation}
|0(\beta )\rangle =\cos \theta |0,\tilde{0}\rangle +\sin \theta |1,\tilde{1}
\rangle .
\label{th}
\end{equation}
where
\begin{eqnarray}
\theta =\tan ^{-1}(e^{-\frac{\beta \omega }{2}}).
\label{angulo}
\end{eqnarray}
The equation (\ref{th}) asserts that the fermionic thermofield vacuum is in the plane generated by $|0,\tilde{0}\rangle $ and $|1,\tilde{1}\rangle $. In fact, the action of the
Bogolioubov transformation on the vacuum excitations (\ref{00}), (\ref{10}), (\ref{01}) and (\ref{11}) is given by 
\begin{equation}
e^{-i\bar{G}}\left( 
\begin{array}{c}
|1,\tilde{1}\rangle  \\ 
|0,\tilde{0}\rangle 
\end{array}%
\right) =\left( 
\begin{array}{cc}
\cos \theta  & -\sin\theta  \\ 
\sin\theta  & \cos \theta 
\end{array}%
\right) \left( 
\begin{array}{c}
|1,\tilde{1}\rangle  \\ 
|0,\tilde{0}\rangle 
\end{array}%
\right)   \label{fermions1}
\end{equation}%
and 
It follows that the fermionic thermofield vacuum, eq. (\ref{th}), is in the plane generated by $|0,\tilde{0}%
\rangle $ and $|1,\tilde{1}\rangle $ and it corresponds to a rotation of $
\theta $, relatively to $|0,\tilde{0}\rangle $. On the other hand, the
action of the Bogolioubov transfomation on (\ref{10}) and (\ref{01}) has no effect, being equivalent to an identity operator in the plane generated by $|1,\tilde{0}%
\rangle $ and $|0,\tilde{1}\rangle $ (see figure \ref{theo12}).
We can use this relation to calculate, for example, the mean value of the number operator
\begin{eqnarray}
\langle \hat{a}^{\dagger }\hat{a}\rangle =\langle 0(\beta )|\hat{a}^{\dagger }\hat{a}|0(\beta
)\rangle =\frac{1}{1+e^{\beta \omega }},
\end{eqnarray}
which is the Fermi-Dirac distribution, where we have agreement with the statistical result as given in the equation (\ref{med}). We can also write \cite{umezawab}
\begin{eqnarray}
\langle \hat{a}^{\dagger }\hat{a}\rangle = \sin^{2}\theta.
\end{eqnarray}

\begin{figure}[h]
\centering
\includegraphics[scale=0.5]{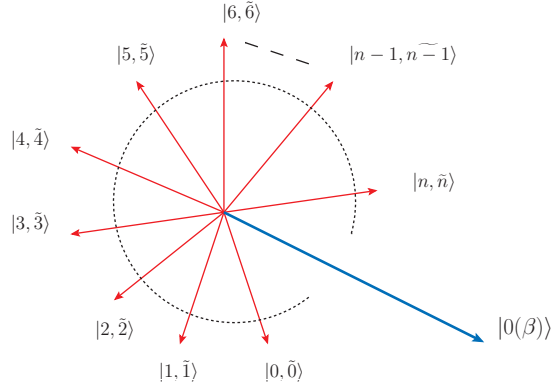}
\caption{(Color online) Bosonic thermofield vacuum in the subspace generated by 
$|0,\tilde{0}\rangle, |1,\tilde{1}\rangle,...,|n,\tilde{n}\rangle, ... \in \mathcal{H}\otimes\tilde{\mathcal{H}}$.}
\label{theo2}
\end{figure}

Similar calculations could be done to the case of a bosonic oscillator, case where the thermofield vacuum is expressed by
\begin{eqnarray}
|0(\beta)\rangle = \sqrt{1-e^{-\beta\omega}}\sum_{n=0}^{\infty}e^{-\frac{n}{2}\beta \omega}|n,\tilde{n}\rangle.
\end{eqnarray}
In this case, the state is generated in the subspace generated by the states 
$|0,\tilde{0}\rangle, |1,\tilde{1}\rangle,...,|n,\tilde{n}\rangle, ... \in \mathcal{H}\otimes\tilde{\mathcal{H}}$, such that 
a representation of the bosonic thermofield vacuum in this subspace is not simple (see figure \ref{theo2}). 
In this case, the state can be described in terms of hiperbolic functions
\begin{eqnarray}
|0(\beta)\rangle &=& \left(\cosh^{-1}\vartheta\right) \sum_{n}\left(\tanh \vartheta \right)^{n}|n,\tilde{n}\rangle,
\end{eqnarray}
where
\begin{eqnarray}
\vartheta=\tanh^{-1}(e^{-\beta\omega/2}).
\end{eqnarray}
As said previously, in terms of a Bogolioubov transformation, we can write $|0(\beta )\rangle =e^{-i\bar{G}}|0,\tilde{0}\rangle$,
where we exchange $\theta$ by $\vartheta$. In this case, mean value of the number operator
\begin{eqnarray}
\langle \hat{a}^{\dagger }\hat{a}\rangle =\langle 0(\beta )|\hat{a}^{\dagger }\hat{a}|0(\beta
)\rangle =\frac{1}{e^{\beta \omega }-1},
\end{eqnarray}
which is the Bose-Einstein distribution. We can also write
\begin{eqnarray}
\langle \hat{a}^{\dagger }\hat{a}\rangle = \sinh^{2}\vartheta.
\end{eqnarray}

\section{Parametric down conversion}

\begin{figure}[h]
\centering
\includegraphics[scale=0.6]{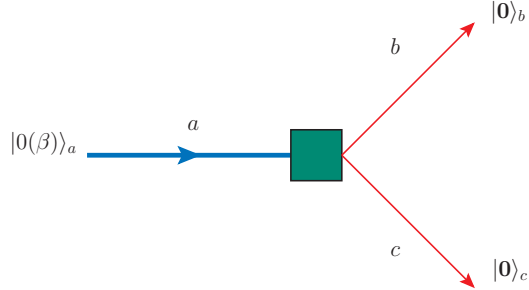}
\caption{ (Color online) Scheme of the parametric down-conversion of the bosonic thermofield vacuum.}
\label{thermo}
\end{figure}

We consider a parametric down conversion described by the following hamiltonian ($\hbar=1$) \cite{vedral}
\begin{eqnarray}
\hat{H}=\hat{a}^{\dagger}\hat{a} + \hat{b}^{\dagger}\hat{b} + \hat{c}^{\dagger}\hat{c} + \kappa \left(\hat{a}\hat{b}^{\dagger}\hat{c}^{\dagger} + \hat{a}^{\dagger}\hat{b}\hat{c} \right).
\end{eqnarray}
In terms of a TFD approach \cite{umezawab,umezawab2,santanab}, a tilde hamiltonian is also constructed
\begin{eqnarray}
\tilde{H}=\tilde{a}^{\dagger}\tilde{a} + \tilde{b}^{\dagger}\tilde{b} + \tilde{c}^{\dagger}\tilde{c} + \kappa \left(\tilde{a}\tilde{b}^{\dagger}\tilde{c}^{\dagger} + \tilde{a}^{\dagger}\tilde{b}\tilde{c} \right).
\end{eqnarray} 
The system evolves according an unitary evolution of the Liouville operator $\bar{L}=\hat{H}-\tilde{H}$, given here by 
\begin{eqnarray}
\bar{L}&=&\omega_{0}\hat{a}^{\dagger}\hat{a}-\omega_{0}\tilde{a}^{\dagger}\tilde{a} 
+ \omega_{1}\hat{b}^{\dagger}\hat{b} -\omega_{1}\tilde{b}^{\dagger}\tilde{b} \nonumber \\ 
&+&  \omega_{2}\hat{c}^{\dagger}\hat{c} -\omega_{2}\tilde{c}^{\dagger}\tilde{c} \nonumber \\ 
&+& \kappa \left(\hat{a}^{\dagger}\hat{b}^{\dagger}\hat{c} - \tilde{a}^{\dagger}\tilde{b}^{\dagger}\tilde{c} 
+ \hat{a}\hat{b}\hat{c}^{\dagger} -\tilde{a}\tilde{b}\tilde{c}^{\dagger} \right)
\end{eqnarray}
Consider the initial state
\begin{eqnarray}
|\psi_{\beta}(0)\rangle = |0(\beta)\rangle_{a}|{\bf0}\rangle_{b}|{\bf0}\rangle_{c}
\end{eqnarray}
where the input state is in the bosonic thermofield vacuum
\begin{eqnarray}
|0(\beta)\rangle_{a} = \sqrt{1-e^{-\beta\omega_{0}}}\sum_{n}e^{-\frac{n}{2}\beta \omega_{0}}|n,\tilde{n}\rangle_{a}
\end{eqnarray}
and the output states are both initially into the vacuum $|{\bf0}\rangle_{b}$ and $|{\bf0}\rangle_{c}$ (see figure \ref{thermo}), 
where
\begin{eqnarray}
|{\bf0}\rangle_{b}&=&|0,\tilde{0}\rangle_{b}\\
|{\bf0}\rangle_{c}&=&|0,\tilde{0}\rangle_{c}.
\end{eqnarray}
By defining
\begin{eqnarray}
\lambda_{n} = e^{-\frac{n}{2}\beta \omega_{0}}\sqrt{1-e^{-\beta\omega_{0}}}
\end{eqnarray}
we can write
\begin{eqnarray}
|\psi_{\beta}(0)\rangle = \sum_{n}\lambda_{n}|n,\tilde{n}\rangle_{a}|0,\tilde{0}\rangle_{b}|0,\tilde{0}\rangle_{c}
\end{eqnarray}
The system evolves according to a Liouville-von Neumann equation
\begin{eqnarray}
\left(\bar{L}-i\partial_{t} \right)|\psi_{\beta}(t)\rangle = 0,
\end{eqnarray}
whose the formal solution is given by
\begin{eqnarray}
|\psi_{\beta}(t)\rangle = e^{-i\bar{L}t}|\psi_{\beta}(0)\rangle.
\end{eqnarray}
We can also write the Liouville operator as
\begin{eqnarray}
\bar{L}&=& \sum_{i=0}^{2}\omega_{i}(\hat{n}_{i} -\tilde{n}_{i}) + \kappa \left(\hat{a}\hat{b}^{\dagger}\hat{c}^{\dagger} - \tilde{a}\tilde{b}^{\dagger}\tilde{c}^{\dagger} + {\texttt{h.c}} \right)
\end{eqnarray}
where the mode frequencies $\omega_{0}$, $\omega_{1}$ and $\omega_{2}$ satisfy
\begin{eqnarray}
\omega_{0}= \omega_{1} + \omega_{2}.
\end{eqnarray}
Under a Liouvillean evolution during a time $t$ and a small coupling term $\lambda$, the initial state
\begin{eqnarray}
|\psi(0)\rangle = |0(\beta)\rangle_{0}|{\bf0}\rangle_{1}|{\bf0}\rangle_{2} \nonumber
\end{eqnarray}
evolves until $\lambda t=1$ leading to the following state
\begin{eqnarray}
|\psi_{\beta}\rangle &=& \frac{1}{\sqrt{2\sum_{n}n\lambda_{n}^{2}}} (-i \sum_{n}\lambda_{n}\sqrt{n}|n-1,\tilde{n}\rangle_{a}|1,\tilde{0}\rangle_{b}|1,\tilde{0}\rangle_{c} 
\nonumber \\
&+& i \sum_{n}\lambda_{n}\sqrt{n}|n,\widetilde{n-1}\rangle_{a}|0,\tilde{1}\rangle_{b}|0,\tilde{1}\rangle_{c} ). 
\label{ufall}
\end{eqnarray}
Initially the thermofield vacuum has a Bose-Einstein distribution given by the mean value of the 
number state $N_{0}=\hat{a}^{\dagger}\hat{a}$, while the expectation value in the other number states 
$N_{1}=\hat{b}^{\dagger}\hat{b}$ and $N_{2}=\hat{c}^{\dagger}\hat{c}$ is zero, 
\begin{eqnarray}
\langle \hat{N}_{0} \rangle_{0} &=& \frac{1}{e^{\beta\omega_{0}}-1}=\sum_{n}n\lambda_{n}^{2}, \\
\langle \hat{N}_{1} \rangle_{0} &=& 0, \\
\langle \hat{N}_{2} \rangle_{0} &=& 0.
\end{eqnarray}
We turn on the interaction time $t$ until the coupling term $\lambda$ and the time satisfy the relation $\lambda t =1$. In this situation, the new expected values
of photon numbers are given by
\begin{eqnarray}
\langle \hat{N}_{0} \rangle &=& \frac{1}{2\langle \hat{N}_{0} \rangle_{0}}\sum_{n}\left(2\lambda_{n}^{2}n^{2} 
-n\lambda_{n}^{2}\right), \\
\langle \hat{N}_{1} \rangle &=& 1,\\
\langle \hat{N}_{2} \rangle &=& 1.
\end{eqnarray}
Then, before the parametric down conversion of the thermofield bosonic vacuum, the mean photon number is increased in each sector measured where there was the states
$|{\bf 0}\rangle_{b}$ and $|{\bf 0}\rangle_{c}$, corresponding to one photon in each side. On the other hand,
the system as a whole is in an entangled state.

The residual distribution on the side $a$, where there was initially a bosonic thermofield vacuum $|0(\beta)\rangle$ can be rewritten
in a more simplified form 
\begin{eqnarray}
\langle \hat{N}_{0} \rangle &=& \langle \hat{N}_{0} \rangle_{0} - 
 \sum_{n\neq m}nm\frac{\lambda_{n}\lambda_{m}}{\langle \hat{N}_{0} \rangle_{0}} -\frac{1}{2}.
\end{eqnarray}
This shows that the thermofield bosonic vacuum lost in photon number in order to create
excitations of $1$-photon states in the sides $b$ and $c$. 

As an example, if we consider $\langle N_{0}\rangle_{0} =2$. 
This corresponds to $T = \omega_{0}\left(\ln(2)\right)^{-1}$ or $\beta = \ln(2)/\omega_{0}$.
It follows that the coefficients $\rho_{n}$ are given by $\rho_{n}=\frac{1}{\sqrt{2}}e^{-\frac{n}{2}\ln(2)}$,
and the final distributon in this case will be given by
\begin{eqnarray}
\langle \hat{N}_{0} \rangle &=& \frac{1}{2}\left[\left(\sum_{n}n^{2}e^{-n\ln(2)}\right) -1\right].
\end{eqnarray}
Considering the bosonic thermofield vacuum until $2$-photon excitations terms, we can write
\begin{eqnarray}
|0(\beta)\rangle_{a} &=& \frac{\sqrt{1-e^{-\beta\omega_{0}}}}{\sqrt{1-e^{-3\beta\omega_{0}}}}( |0,\tilde{0}\rangle_{a} 
+ e^{-\frac{1}{2}\beta \omega_{0}}|1,\tilde{1}\rangle_{a} \nonumber \\
&+& e^{-\beta \omega_{0}}|2,\tilde{2}\rangle_{a} )
\end{eqnarray}
and the state is now written as
\begin{eqnarray}
|\psi_{\beta}(t)\rangle &=& \frac{1}{\sqrt{2\lambda_{1}^{2}+4\lambda_{2}^{2}}} ( 
-i\lambda_{1}|0,\tilde{1}\rangle_{a}|1,\tilde{0}\rangle_{b}|1,\tilde{0}\rangle_{c} \nonumber \\
&-& i\lambda_{2}\sqrt{2}|1,\tilde{2}\rangle_{a}|1,\tilde{0}\rangle_{b}|1,\tilde{0}\rangle_{c}  
\nonumber \\
&+& i \left[\lambda_{1}|1,\tilde{0}\rangle_{a} + \lambda_{2}\sqrt{2}|2,\tilde{1}\rangle_{a} \right]|0,\tilde{1}\rangle_{b}|0,\tilde{1}\rangle_{c}) 
\nonumber \\
\end{eqnarray}
If we project the side $a$ in the state $|0,\tilde{0}\rangle_{a}$, we have
\begin{eqnarray}
_{a}\langle 0,\tilde{0}|\psi_{\beta}(t)\rangle &=& 0.
\end{eqnarray}
This implies that this state has no parts in the vacuum and then cannot be projected into it. The states have
evolved such that the projection into the one-photon excitations will lead
\begin{eqnarray}
_{a}\langle 0,\tilde{1}|\psi_{\beta}(t)\rangle &=& \frac{-i\lambda_{1}}{\sqrt{2\lambda_{1}^{2}+4\lambda_{2}^{2}}}|1,\tilde{0}\rangle_{b}|1,\tilde{0}\rangle_{c} \nonumber \\
\end{eqnarray}
\begin{eqnarray}
_{a}\langle 1,\tilde{0}|\psi_{\beta}(t)\rangle &=& 
\frac{i\lambda_{1}}{\sqrt{2\lambda_{1}^{2}+4\lambda_{2}^{2}}}|0,\tilde{1}\rangle_{b}|0,\tilde{1}\rangle_{c} 
\nonumber \\
\end{eqnarray}
\begin{eqnarray}
_{a}\langle 2,\tilde{1}|\psi_{\beta}(t)\rangle &=&\frac{i\lambda_{2}\sqrt{2}}{\sqrt{2\lambda_{1}^{2}+4\lambda_{2}^{2}}}|0,\tilde{1}\rangle_{b}|0,\tilde{1}\rangle_{c}
\end{eqnarray}
\begin{eqnarray}
_{a}\langle 1,\tilde{2}|\psi_{\beta}(t)\rangle &=&\frac{-i\lambda_{2}\sqrt{2}}{\sqrt{2\lambda_{1}^{2}+4\lambda_{2}^{2}}}|1,\tilde{0}\rangle_{b}|1,\tilde{0}\rangle_{c}
\end{eqnarray}
These projections means that all the possible measured values in $a$ will lead to states of one-photon excited states in 
$b$ and $c$, and these states are completelly separable in the space $\mathcal{H}\otimes\tilde{\mathcal{H}}$. This result
is better to explain the previous one (\ref{ufall}). In fact, the projections 
\begin{eqnarray}
_{a}\langle n-1,\tilde{n}|\psi_{\beta}\rangle &=& \frac{-i \lambda_{n}\sqrt{n}}{\sqrt{2\sum_{n}n\lambda_{n}^{2}}}|1,\tilde{0}\rangle_{b}|1,\tilde{0}\rangle_{c} 
\end{eqnarray}
and
\begin{eqnarray}
_{a}\langle n,\widetilde{n-1}|\psi_{\beta}\rangle &=& \frac{i\lambda_{n}\sqrt{n}}{\sqrt{2\sum_{n}n\lambda_{n}^{2}}}|0,\tilde{1}\rangle_{b}|0,\tilde{1}\rangle_{c}
\end{eqnarray}
Consequently, in the approximation considered the parametric down conversion of thermofield vacuum will lead to only
excitations of one-photon states. Once specified the state of a Fock state $\mathcal{H}\otimes\tilde{\mathcal{H}}$
 in $a$, the states in $b$ and $c$ are specified as excitations of one photon states in $\mathcal{H}\otimes\tilde{\mathcal{H}}$
 and are not entangled in the space $\mathcal{H}\otimes\tilde{\mathcal{H}}$.  It follows that a set of measurements of the residual state in the Fock basis of $\mathcal{H}\otimes\tilde{\mathcal{H}}$ will determine
clearly the excitations of one-photon distributions in the sides $a$ and $b$.

\section{Conclusion}

We have considered a parametric down conversion where the input state is given by a thermofield bosonic vacuum 
state on a side $a$, related to a Bose-Einstein distribution. Turning on the interaction until time of interaction and the coupling constant
satisfy $\lambda t=1$ and taking into account a small coupling term $\lambda$, we arrive at the result where the output
states on sides $b$ and $c$, initially on vacuum states, are excited in one-photon states, leading
to number operator expectation values of one photon
number in each side, $b$ and $c$.

On the other hand, the thermofield bosonic vacuum state is not totally annihilated in the process, resulting in a residual
state whose profile of the number operator expectation value is different 
from a Bose-Einstein distribution, due to losing of photons in the interaction. 

We shown that a specific measurements on the Fock state of the space $\mathcal{H}\otimes\tilde{\mathcal{H}}$ will lead
to a specification of one-photon states distributions in the sides $b$ and $c$. 

Such ideas can also be of relevance on the experimental side \cite{bergeal}, where thermofield or thermal like states
are involved.

\section{Acknowledgements}

The author thanks CAPES (Brazil) for financial support.

\end{document}